\documentclass[12pt]{article}
\usepackage[top=2.0cm,bottom=2.0cm,left=2.54cm,right=2.54cm]{geometry}
\usepackage{
amsmath,
amssymb,
graphicx,subfigure,caption,
comment,
booktabs,
cite}
\usepackage{bm}
\usepackage{color}  
\usepackage{extarrows}

\usepackage[hyperindex=true,
          pdfstartview=FitH,
          bookmarksnumbered=true,
          bookmarksopen=true,
          citecolor=blue,
          linkcolor=blue,
          colorlinks=true,
          pdfborder=001,
          unicode]{hyperref}

\allowdisplaybreaks

\newcommand{\D}{{\rm d}}

\renewcommand\({\left(}
\renewcommand\){\right)}

\newcommand{\blue}[1]{\textcolor{blue}{#1}}
\renewcommand{\blue}[1]{\textcolor{black}{#1}}

\newcommand{\ind}[2]{^{#1}{}_{#2}}

\newcommand{\add}{\text{add}}
\newcommand{\fa}{\mathfrak{a}}
\newcommand{\fb}{\mathfrak{b}}
\newcommand{\fd}{\mathfrak{d}}

\newcommand{\re}{{\rm e}}
\newcommand{\pfrac}[2]{\frac{\partial #1}{\partial #2}}

\begin{document}
\title{Relativistic stochastic mechanics I:\\
Langevin equation from observer's perspective}

\author{
Yifan Cai\thanks{{\em email}: \href{mailto:caiyifan@mail.nankai.edu.cn}
{caiyifan@mail.nankai.edu.cn}},~
Tao Wang\thanks{
{\em email}: \href{mailto:taowang@mail.nankai.edu.cn}{taowang@mail.nankai.edu.cn}},~
and Liu Zhao\thanks{Corresponding author, {\em email}: 
\href{mailto:lzhao@nankai.edu.cn}{lzhao@nankai.edu.cn}}\\
School of Physics, Nankai University, Tianjin 300071, China}

\date{}
\maketitle
\begin{abstract}
Two different versions of relativistic Langevin equation in curved spacetime
background are constructed, both are 
manifestly general covariant. It is argued that, from the observer's point of view, 
the version which takes the proper time of the Brownian particle as evolution 
parameter contains some conceptual issues, while the one which makes use of
the proper time of the observer is more physically sound. The two versions of 
the relativistic Langevin equation are connected by a reparametrization scheme. 
In spite of the issues contained in the first version of the 
relativistic Langevin equation, it still permits to extract the physical 
probability distributions of the Brownian particles, as is shown by Monte Carlo 
simulation in the example case of Brownian motion in $(1+1)$-dimensional
Minkowski spacetime.
\end{abstract}

\section{Introduction}
\label{introduction}

General relativity and non-equilibrium statistical physics are two 
important frontiers of modern theoretical physics. In spite of the significant
progresses in their respective fields, the study on the overlap between these 
two fields remains inactive. However, owing to the development in astrophysics, 
there are more and more scenarios in which both general relativity and 
non-equilibrium statistical physics are important. Therefore, it becomes 
necessary and of utmost importance to take the combination of general relativity and 
non-equilibrium statistical physics more seriously. 

There are two major branches in non-relativistic non-equilibrium statistical 
physics, i.e. kinetic theory and stochastic mechanics. The study of kinetic theory 
started from Boltzmann's works, and its relativistic version also has a long history
(which can be traced back to J\"uttner's works in 1911 \cite{Juttner1911}). Currently, 
the framework of relativistic kinetic theory looks fairly complete 
\cite{degroot1980relativistic,cercignani2002relativistic,
vereshchagin2017relativistic, acuna2022introduction}.
In contrast, the study of relativistic stochastic mechanics is still far from being
accomplished. \blue{Since the relativistic Ornstein-Uhlenbeck process was proposed 
about 20 years ago\cite{debbasch1997relativistic}}, there appeared some attempts in 
relativistic stochastic mechanics \cite{debbasch2004diffusion,
dunkel2005theory1,dunkel2005theory2,fingerle2007relativistic,franchi2007relativistic,
dunkel2009relativistic,dunkel2009time,herrmann2009diffusion,herrmann2010diffusion}, 
mostly in the special relativistic regime. However, \blue{apart from Herrmann 
\cite{herrmann2009diffusion,herrmann2010diffusion} and Haba's \cite{haba2010relativistic}
works,} the manifest covariance of stochastic mechanics is typically absent. 
\blue{Some work \cite{franchi2007relativistic} considered concrete curved spacetime 
background without paying particular attention to general covariance. 
There are also some other works which focus on the covariance of 
stochastic thermodynamics \cite{ding2020covariant,
ding2022covariant}, but those works have nothing to do with relativity.}

The random motion of heavy particles began to attract scientific interests in the 
late 19th and early 20th centuries, as it provides a simple example for the 
diffusion phenomena. Einstein \cite{einstein1905neue,einstein1905molekularkinetischen} 
and Smoluchowski \cite{smoluchowski1906kinetischen} showed that the random motion 
is closely related to macroscopic environment, however, the microscopic description 
of the random motion has not been established. Later, Langevin \cite{langevin1908theorie} 
wrote down the first equation of motion for a Brownian particle by his physical 
intuition, which inspired subsequent explorations about the microscopic mechanisms of 
Brownian motion. In the 1960-1970s, a series of models  
\cite{ford1965statistical, mori1965transport, zwanzig1973nonlinear} were 
proposed in this direction, which made it clear why the disturbance from the 
heat reservoir could be viewed as Gaussian noises, and hence a bridge between 
microscopic mechanical laws and non-equilibrium macroscopic phenomena is 
preliminarily established in the non-relativistic regime. Since the 1990s,  
the so-called stochastic thermodynamics based on top of Langevin equation 
was established \cite{sekimoto1998langevin,sekimoto2010stochastic}. 

To some extent, the challenge in constructing a covariant Langevin equation 
arises from the underestimation about the role of the observer. Unlike general 
relativity which concentrates mainly on the universal observer independent laws about 
the spacetime, statistical physics concentrates more on the observational or
phenomenological aspects, which are doomed to be observer dependent. The lack of 
manifest covariance in some of the works on relativistic Langevin equation, e.g. 
\cite{debbasch1997relativistic,debbasch2004diffusion,dunkel2005theory1,
dunkel2005theory2,fingerle2007relativistic}, stems from the choice of the 
coordinate time as evolution parameter. As exceptional examples, 
\blue{Herrmann \cite{herrmann2009diffusion,herrmann2010diffusion} 
and Haba's \cite{haba2010relativistic} work }adopted the proper time 
of the Brownian particle evolution parameter and the corresponding versions 
of Langevin equation are indeed manifestly covariant. Nevertheless, the role of the 
observer is still not sufficiently stressed in those works, and
it will be clear that, from the observer's point of view, 
the proper time of the Brownian particle should not be thought of as an 
appropriate evolution parameter. The present work 
aims to improve the situation by reformulating the relativistic  Langevin equation
from the observer's perspective and taking the observer's proper time as evolution 
parameter. In this way, we obtain the general relativistic
Langevin equation which is both manifestly general covariant and explicitly 
observer dependent. 

This work is Part I of a series of two papers under the same main 
title ``Relativistic stochastic mechanics''. Part II will be concentrated on the 
construction of Fokker-Planck equations associated with the Langevin equations 
presented here.

This paper is organized as follows. In Sec.\ref{Observers_and_State_Space}, 
we first clarify certain conceptual aspects of relativistic mechanics which are otherwise 
absent in the non-relativistic context. These include the explanation on the role of 
observers, the choice of time and the conventions on the space of micro states (SoMS). 
Sec.\ref{The_Covariant_Langevin_Equation} is devoted to a first attempt 
for the construction of general relativistic Langevin equation. To make the 
discussions self contained, we start from a brief review about the non-relativistic 
Langevin equation, and then pay special attentions toward the form of the damping 
and additional stochastic forces in the relativistic regime. 
As the outcome of these analysis, we write down a 
first candidate for the relativistic Langevin equation [referred to as LE$_\tau$], which is 
manifestly general covariant. It is checked that the stochastic motion of the 
Brownian particle following this version of the Langevin equation does not 
break the mass shell condition. However, since this version of the 
relativistic Langevin equation employs the proper time $\tau$ of the 
Brownian particle as evolution parameter, there are still
some issues involved in it, because, from the observer's point of view, $\tau$ 
itself is a random variable, which is inappropriate to be taken as an evolution 
parameter. The problem with LE$_\tau$ is resolved in Sec.\ref{Reparametrization} 
by introducing a reparametrization scheme, which yields another version 
of the relativistic Langevin equation [LE$_t$], which employs the proper time $t$ of the 
prescribed observer as evolution parameter. In Sec.\ref{MonteCarlo}, 
the stochastic motion of Brownian particles in $(1+1)$-dimensional Minkowski spacetime 
driven by a single Wiener process and subjects to an isotropic homogeneous 
damping force is analyzed by means of Monte Carlo simulation. It is shown that, 
in spite of the issues mentioned above, LE$_\tau$ still permits for exploring the  
physical probability distributions, and the resulting distributions 
are basically identical to those obtained from LE$_t$. Finally, we present some
brief concluding remarks in Sec.\ref{Conclusion}.

\section{Observers, time, and the SoMS}
\label{Observers_and_State_Space}

As mentioned earlier, we are interested in describing the stochastic motion of Brownian
particles in a generic spacetime manifold $\mathcal M$. To achieve this goal, a fully 
general covariant description for the SoMS and equations of motion 
are essential.

Determining a micro state of a classical physical system  
requires the simultaneous determination of the concrete position and momentum of 
each individual particle at a given instance of time. In Newtonian mechanics, 
there is an absolute time, therefore, there is no ambiguity as to what constitutes 
a ``given instance of time''. However, in relativistic regime, the concept of 
simultaneity becomes relative, and in order to assign a proper meaning for a 
micro state, one needs to introduce a concrete time slicing (or temporal foliation) 
of the spacetime at first. There are two approaches to do so, i.e. (1) choosing some 
coordinate system and making use of the coordinate time as the slicing parameter; 
(2) introducing some properly aligned observer field and choosing the proper velocity 
$Z^\mu$ ($Z^\mu Z_\mu=-1$) of the observer field as normalized normal vector field of 
the spatial hypersurfaces consisting of ``simultaneous events'', which is 
also referred to as the configuration space. Let us recall that an observer 
in a generic $(d+1)$-dimensional spacetime manifold $\mathcal M$ is 
represented by a timelike curve with normalized 
future-directed tangent vector $Z^\mu$ which is identified as the proper velocity 
of the observer. An observer field is a densely populated collection of observers
whose worldlines span the full spacetime. The second slicing approach is always 
possible because each observer naturally carries a Frenet frame with 
orthonormal basis $e^\mu{}_{\hat \nu}$ with $e\ind{\mu}{\hat 0}:=Z^\mu$, and, 
as one of the basis vector field, $Z^\mu$ naturally satisfies the Frobenius theorem
\begin{align*}
    Z_{[\mu}\nabla_\nu Z_{\rho]}=0,
\end{align*}
which in turn implies the existence of spacelike hypersurfaces which take $Z^\mu$ 
as normal vector field.

In practice, the two time slicing approaches can be made identical. One only needs to 
choose the specific observers whose proper velocity covector 
field $Z_\mu$ is proportional to $(\D {x^0})_\mu$. However, such an identification 
often obscures the role of the observer field and brings about the illusion that 
the corresponding description is necessarily coordinate dependent 
and lacks the spacetime covariance. Therefore, it will be preferable to
take the choice of not binding the coordinate system and the observer field 
together and focusing on explicit general covariance.

While an observer field can be used to identify which events happens simultaneously, 
it cannot uniquely specify the timing of the configuration spaces. To achieve this, 
we need to pick a single observer, referred to as {\em Alice}, from the set of observers. 
The integral curve of this particular observer can be denoted 
as $x^\mu(t)$, where $t$ represents the proper time of 
this single observer. In principle, we can extend $t$ into a smooth scalar field $t(x)$ 
over the whole spacetime manifold, such that we can label the configuration space 
at the proper time $t$ of Alice unambiguously as the hypersurface 
$\mathcal S_t:=\{x\in {\mathcal M}| t(x)=t={\rm const}.\}$. 
The union of $\mathcal S_t$ at all possible $t$
covers $\mathcal M$. Notice that, in general,  $t$, $x^0$ (the zeroth 
component of the coordinate system) and $\tau$ (the proper time of the Brownian
particle) can all be different entities. 

The momentum of a relativistic particle is a tangent vector of the 
spacetime manifold \footnote{Since the space time is assumed to be 
endowed with a non-degenerate metric $g_{\mu\nu}(x)$, we can identify the cotangent 
vector at any event with its dual tangent vector. Therefore, we are free to take 
the tangent space description instead of the cotangent space description 
in this work.}
$\mathcal M$. Accordingly, the momentum space of a particle should be a subset of the 
tangent bundle $T\mathcal M$ of the spacetime, because the momentum must obey the 
mass shell condition 
\begin{align}
S(x,p):= p^\mu p_\mu +m^2 = g_{\mu\nu}(x) p^{\mu} p^{\nu} + m^{2}=0.  
\label{massshell}
\end{align}
Moreover, the momentum 
of a massive particle must be a future-directed timelike vector, i.e.
$p^{\mu} Z_{\mu}(x)<0$. Putting these requirements together, we conclude that 
the SoMS of a massive relativistic particle must be a subspace of 
the future mass shell bundle $\Gamma^{+}_{m}$,
\begin{align*}
\Gamma^{+}_{m}:=\{(x,p)\in T\mathcal M|\ g_{\mu\nu}(x) p^{\mu} p^{\nu} 
=-m^{2}\ \text{and} \ p^{\mu} Z_{\mu}(x)<0\}.
\end{align*}
\blue{The geometry of future mass shell bundle is decided by the Sasaki metric\cite{sarbach2014geo}, and its associated volume element is}
\begin{align}\label{volume-element}
	\eta_{\Gamma^{+}_{m}}&=\frac{\det(g)}{p_{0}}\D x^{0}\wedge\D x^{1}\wedge...\wedge\D x^{d}\wedge\D p^{1}\wedge...\wedge \D p^{d}.
\end{align}
The momentum space at the event $x\in\mathcal M$ is simply the fiber of the
future mass shell bundle $\Gamma^{+}_{m}$ at the base point $x$,
\begin{align*}
	(\Gamma^{+}_{m})_{x}:=T_{x} \mathcal M\cap\Gamma^{+}_{m}.
\end{align*}
Please be aware that the SoMS of a massive relativistic particle 
is {\em not} the full future mass shell bundle $\Gamma^{+}_{m}$, 
because the configuration space is only the 
spacelike hypersurface $\mathcal S_t$ consisted of simultaneous events 
regarding to the proper time $t$ of Alice. Therefore, the actual 
SoMS should be the proper subspace 
\begin{align*}
    \Sigma_t:=\bigcup_{x\in\mathcal S_t}(\Gamma_m^+)_x
    =\{(x,p)\in \Gamma^+_m| x\in\mathcal S_t \}
\end{align*}
of the full future mass shell bundle $\Gamma^{+}_{m}$.

Due to the mass shell condition, the actual momentum space $(\Gamma^{+}_{m})_{x}$ 
has one less dimension than the tangent space $T_x \mathcal M$. It will be appropriate to 
think of $(\Gamma^{+}_{m})_{x}$ as a codimension one hypersurface in 
$T_x \mathcal M$ defined via eq.\eqref{massshell}, and its (unmormalized) normal 
covector can be defined as $\mathcal N_\mu:= \frac{\partial}{
\partial p^\mu} S(x,p)=2p_\mu$. In view of this, any tangent vector field 
$\mathcal V\in T[(\Gamma^{+}_{m})_{x}]$ must be normal to $p_\mu$, 
i.e., $\mathcal V^\mu p_\mu=0$. Using this property, we can select a basis for 
the tangent space of the momentum space, i.e.
\begin{align*}
\dfrac{\partial}{\partial \breve{p}^{i}}
:=\dfrac{\partial}{\partial {p}^{i}}   
-\dfrac{p_{i}}{p_{0}}\dfrac{\partial}{\partial {p}^{0}}.
\end{align*}
Therefore, the tangent vector field $\mathcal V\in T[(\Gamma^{+}_{m})_{x}]$ acquires 
two component-representations, one in the basis 
$\dfrac{\partial}{\partial p^{\mu}}$ of 
$T(T_x \mathcal M)$, and one in the basis $\dfrac{\partial}{\partial \breve{p}^{i}}$ of 
$T[(\Gamma^{+}_{m})_{x}]$. It is easy to check that these two representations 
are equivalent, 
\begin{align}\label{relation-vector-on-mass-shell}
\mathcal{V}^{i}\frac{\partial}{\partial \breve{p}^{i}}  
=\mathcal{V}^{i}\frac{\partial}{\partial p^{i}}   
-\mathcal{V}^{i}\frac{p_{i}}{p_{0}}\frac{\partial}{\partial p^{0}}   
=\mathcal{V}^{i}\frac{\partial}{\partial p^{i}}   
+\mathcal{V}^{0}\frac{\partial}{\partial p^{0}}  
 =\mathcal{V}^{\mu}\dfrac{\partial}{\partial p^{\mu}}.
\end{align} 
Therefore, when describing a vector in $T[(\Gamma^{+}_{m})_{x}]$, the two 
component-representations $\mathcal{V}^{i}$ and $\mathcal{V}^{\mu}$ can be 
used interchangeably.

\section{The covariant Langevin equation}
\label{The_Covariant_Langevin_Equation}

\subsection{A short review of Langevin equation in non-relativistic setting}

The main focus of this section is to construct a covariant Langevin equation in a 
generic spacetime. Before dwelling into the detailed construction, it seems helpful 
to make a brief review of Langevin equation in the non-relativistic setting. 

The non-relativistic Langevin equation describes the motion of a Brownian particle 
in a fixed heat reservoir. The initial intuitive construction of Langevin equation 
is simply based on the second law of Newtonian mechanics, in which the motion of the
Brownian particle is driven by drift and damping forces $F_{\rm drift}(x)$, 
$F_{\rm damp}(p)$ together with a random force $\xi(t)$. The drift force $F_{\rm drift}(x)$
is provided by a conservative potential and hence is dependent on the coordinate 
position $x$ of the Brownian particle. The damping force $F_{\rm damp}(p)$, however, 
is dependent on the momentum of the particle. 
In the absence of the drift force, the Langevin equation describing one-dimensional 
Brownian motion can be intuitively written as
\begin{align}
\dfrac{\D p}{\D t} = F_{\rm damp}(p) + \xi(t).  \label{intuitiveLE}
\end{align}
However, it was soon realized that this intuitive picture cannot be mathematically correct,
because, under the impact of the random force, the momentum of the Brownian particle 
cannot be differentiable with respect to the time $t$, and hence the Langevin equation 
cannot actually be regarded as a differential equation. 

The modern understanding of Langevin equation is as follows. Consider a scenario 
in which a large number of light particles exist in the heat reservoir, and they randomly 
collide with the heavy Brownian particle, causing the momentum of the latter to alter with 
each collision. If the mass ratio between the Brownian particle and the particle from the 
heat reservoir is sufficiently large, there will be a timescale $\D t$ during which 
a sufficiently large number of independent collisions happen. Since the Brownian particle 
is heavy, its state changes very little within this timescale. 
According to the central limit theorem, the probability 
distribution of the variations of the momentum during $\D t$ follows a 
Gaussian distribution. 
The average value of this distribution yields the damping force $F_{\rm damp}$, 
while the remaining (rapid) portion is viewed as a stochastic force. Thus, 
the classical Langevin equation in one-dimensional space can be expressed as
\begin{align}
    \D \tilde x_n &= \frac{\tilde p_n}{m}\D t\notag\\
    \D \tilde p_n &= R\D\tilde w_n + F_{\rm damp} \D t, 
    \label{disLE}
\end{align}
where the suffices $n$ represents the $n$-th time step and 
$\D\tilde w_n$ is a random variable obeying Gaussian distribution 
\begin{align*}
    \Pr[\D\tilde w_n = \D w_n] 
    = \frac{1}{\sqrt{2\pi\D t}}\re^{-\frac{1}{2}\frac{\D w_n^2}{\D t}}.
\end{align*}
The coefficient $R$ appeared in eq.\eqref{disLE} is called stochastic amplitude. 
Notice that the variance of the above Gaussian distribution equals $\D t$.
In this paper, tilded variables such as $\tilde x, \tilde p$ 
represent random variables, and the corresponding un-tilded symbols (e.g. $x,p$) 
represent their concrete realizations. 

The Langevin equation as presented above is technically a system of discrete-time difference 
equations, as the time scale $\D t$ must be large enough to permit sufficient 
number of collisions to happen during this time interval. However, if $\D t$ is far smaller 
than the relaxation time, it can be effectively thought of as infinitesimal. 
In the limit of continuity, $\tilde w_n$ becomes a Wiener process $\tilde w_t$, 
and there is an ambiguity in the coupling rule between the stochastic amplitude $R$ 
and the Wiener process if $R$ is dependent on the momentum. Unlike in normal calculus, 
in the continuity limit, 
\begin{align}
     R(\tilde p_n+a\D\tilde p_n)\D \tilde w_n
    \xlongrightarrow[]{\D t\rightarrow 0} R(\tilde p_t)\circ_a\D \tilde w_t 
    \label{arule}
\end{align}
depends on the value of $a\in[0,1]$ \cite{gardiner1985handbook}. The continuum 
version of Langevin equation with the above coupling rule reads
\begin{align}
    \D \tilde x_t &= \frac{\tilde p_t}{m}\D t\notag\\
    \D \tilde p_t &= R(\tilde p_t) \circ_a \D\tilde w_t + F_{\rm damp} \D t .
    \label{LEcirca}
\end{align}
In particular, the coupling rule with $a=0$ is known as 
Ito's rule and is denoted as $R\circ_I\D \tilde w_t$, while the rule with $a=1/2$ is 
known as Stratonovich's rule and is denoted as $R\circ_S\D \tilde w_t$. 

Ito's rule allows Langevin equation to be understood as an equation describing a 
Markov process, making it easier to analyze. However, since $\D t$ is equal 
to the variance of the Wiener process, $\D \tilde w_t$ should be in the same order of 
magnitude with $\sqrt{\D t}$. This fact leads to some profound consequences. 
For instance, it can be easily verified that any coupling rule $\circ_a$ can be related to 
Ito's rule via
\begin{align*}
R(\tilde p_t)\circ_a\D \tilde w_t 
= R(\tilde p_t)\circ_I\D \tilde w_t +a R \frac{\partial}{\partial p}R\,\D t,
\end{align*}
which is a straightforward consequence of eq.\eqref{arule}. Moreover, 
it can also be checked that Ito's rule breaks the chain rule of calculus,
\begin{align*}
\D h(\tilde p_t)   &=\frac{\partial h}{\partial p}\D \tilde p_t
+\frac{1}{2}\frac{\partial^2 h}{\partial p^2}\D \tilde p_t^2\notag\\   
&=\left(\frac{\partial h}{\partial p}F_{\rm damp}
+\frac{1}{2}\frac{\partial^2 h}{\partial p^2}R^2\right) \D t 
+ \frac{\partial h}{\partial p}R(\tilde p_t)\circ_I\D \tilde w_t
\neq \frac{\partial h}{\partial p} \D \tilde p_t.
\end{align*}
On the other hand, Stratonovich's rule is the unique rule that preserves the chain rule, 
\begin{align*}
\D h(\tilde p_t)
&=\frac{\partial h}{\partial p}\D \tilde p_t
+\frac{1}{2}\frac{\partial^2 h}{\partial p^2}\D \tilde p_t^2\notag\\
&= \frac{\partial h}{\partial p}
\(R(\tilde p_t) \circ_a \D\tilde w_t + F_{\rm damp} \D t\)
+\frac{1}{2}\frac{\partial^2 h}{\partial p^2}
\(R(\tilde p_t) \circ_a \D\tilde w_t + F_{\rm damp} \D t\)^2
\notag\\
&=\(\frac{\partial h}{\partial p}F_{\rm damp}
+a\frac{\partial h}{\partial p}R\frac{\partial}{\partial p}R
+\frac{1}{2}\frac{\partial^2 h}{\partial p^2}R^2\)\D t
+\frac{\partial h}{\partial p}R(\tilde p_t)\circ_I\D \tilde w_t\notag\\
&\xlongequal[]{a={1}/{2}} \frac{\partial h}{\partial p}F_{\rm damp}\D t
+\frac{1}{2}\frac{\partial h}{\partial p}R\frac{\partial}{\partial h}
\( \frac{\partial h}{\partial p}R \) \D t
+\frac{\partial h}{\partial p}R(\tilde p_t)\circ_I\D \tilde w_t\notag\\
&=\frac{\partial h}{\partial p}F_{\rm damp}\D t
+\frac{\partial h}{\partial p}R\circ_S\D \tilde w_t
=\frac{\partial h}{\partial p} \D \tilde p_t,
\end{align*}
where, in the last step, we used the Langevin equation which adopts Stratonovich's rule. 
Since the tensor calculus on manifolds is strongly 
dependent on the chain rule, it is natural to adopt Stratonovich's rule while constructing 
the general covariant Langevin equation on a generic spacetime manifold, 
as we will do in the subsequent analysis. Other
elaborations on the covariance of Stratonovich type stochastic differential 
equations can be found in Refs. \cite{hsu2002stochastic, armstrong2016coordinate,
armstrong2018intrinsic}.

\subsection{Nonlinear damping force and additional stochastic force}

Let us now consider the Langevin equation \eqref{LEcirca} with Ito's rule and 
make a comparison with the intuitive form \eqref{intuitiveLE} of the equation. 
In essence, both the damping force $F_{\rm damp}$ and ``stochastic force'' 
$\xi(t)=R\circ_I\D\tilde w/\D t$ arise from the collisions between the Brownian particle 
and the heat reservoir particles, however, we have artificially separated them.  
It is possible to derive the expressions of $R$ and $F_{\rm damp}$ directly from 
microscopic mechanics and the chaotic assumption of the heat reservoir 
\cite{mori1965transport,zwanzig1973nonlinear}. Macroscopically, 
the stochastic force can be viewed as the consequence of thermal fluctuations, and thus 
it vanishes in the low temperature limit. In such surroundings, the damping force 
can be measured directly, allowing us to construct simple phenomenological models. 
The simplest one assumes a linear damping force proportional to the momentum of 
the Brownian particle in the reference frame comoving with the heat reservoir:
\begin{align*}
    F_{\rm damp}=-Kp,
\end{align*}
where $K$ is the damping coefficient. 
This simple model captures two important features of the damping force: First, when 
the Brownian particle comoves with the heat reservoir, the damping force vanishes. 
Second, the direction of the damping force should be opposite to the relative 
velocity. In more general cases than the linear  
damping model, the damping coefficient $K$ could be dependent on the momentum $p$ 
of the Brownian particle. 

If $K$ is independent of $p$, the stochastic amplitude 
$R$ can be easily derived using the thermal equilibrium between the Brownian particle 
and the heat reservoir, yielding
\begin{align*}
D:=R^2=2\,TK,
\end{align*}
where $T$ is the temperature of the reservoir. 
This relation is known as the Einstein relation. However, for nonlinear 
damping forces, the situations become much more complicated. 
Ref.\cite{klimontovich1994nonlinear} demonstrated that, provided $R$ is momentum dependent, 
there exists a non-zero force $\frac{1}{2}{\partial (R^2)}/{\partial p}$ 
acting on the Brownian particle even if its momentum vanishes. This extra force term   
is also a consequence of the thermal equilibrium between the Brownian 
particle and the heat reservoir. There are two options for interpreting this extra force. 
The first option is to consider it as a part of the damping force, so that the 
full damping force takes the form
\begin{align*}
    F_{\rm damp}=R\frac{\partial}{\partial p}R-K p =-K_{\text{eff}} \,p,
\end{align*}
where the effective damping coefficient reads 
\begin{align*}
    K_{\text{eff}}:=K-\frac{R}{p}\frac{\partial}{\partial p}R.
\end{align*}
Consequently, there will be a modified Einstein relation
\begin{align*}
    R^2=2T\left[K_{\text{eff}}+\frac{R}{p}\frac{\partial}{\partial p}R\right].
\end{align*}
This option seems to have several issues. (1) It looks strange that the 
damping force still exists when the momentum is zero; (2) 
More importantly, we cannot define an effective damping coefficient  
in higher spatial dimensions using this approach. The second option is to split 
the extra force term $\frac{1}{2} {\partial R^2}/{\partial p}$ into two equal 
halves: one half is to be combined with the Ito's coupling 
to give rise to Stratonovich's coupling with Gaussian noises, 
and the other half is understood as an ``additional stochastic force''
\begin{align*}
F_{\add}:=\frac{1}{2}R\frac{\partial}{\partial p}R.
\end{align*} 
Hence, the more general Langevin equation in $d$-dimensional flat space can be written as
\begin{align*}
    \D \tilde x^i_t&=\frac{\tilde p^i_t}{m}\D t\notag\\
    \D\tilde p^i_t &=\left[ R\ind{i}{\fa}\circ_S\D\tilde w_t^\fa
    +F^i_{\add}\D t \right] -K\ind{i}{j}\tilde p^j_t\D t,
\end{align*}
where the indices $i, j$ label different spatial dimensions and $\fa,\fb$ 
are used to distinguish independent Gaussian noises. It should be remarked that 
the number $\fd$ of Gaussian noises is independent of the dimension $d$ of the space. 
The additional stochastic force now reads
\begin{align}
F^i_\add=\frac{\delta^{\fa\fb}}{2}R\ind{i}{\fa}\frac{\partial}{\partial p^j}R\ind{j}{\fb}.
\label{faddnonr}
\end{align}

The discussions made so far in this subsection have been restricted to the non-relativistic
situations. In the next subsection, it will be clear that the mass shell condition in 
the relativistic setting requires that the damping coefficients have to be 
momentum dependent. Therefore, the additional stochastic force should also 
appear in the relativistic Langevin equation. To derive the concrete expression 
for this additional stochastic force, we need to make use of the Fokker-Planck 
equation and the relativistic Einstein relation. However, since the focus of the 
present work is on the relativistic Langevin equation, we will provide the detailed 
derivation in Part II of this series of works. At present, we simply provide the result,
\begin{align}\label{additional}
\mathcal F^\mu_\add 
=\frac{\delta^{\fa\fb}}{2}\mathcal R\ind{\mu}{\fa}\nabla^{(h)}_i\mathcal R\ind{i}{\fb},
\end{align}
where $h$ refers to the metric on the mass shell $(\Gamma^+_m)_x$.
It is important to note that the stochastic amplitudes $\mathcal R\ind{ \mu }{ \fa }$ 
in the relativistic Langevin equation should be a set of vectors on the curved 
Riemannian manifold $(\Gamma^+_m)_x$, i.e. $\mathcal R\ind{ \mu }{ \fa } \in 
T[(\Gamma^+_m)_x]$. As such, the derivative operator $\partial/\partial p^i$ 
that appeared in eq.\eqref{faddnonr} needs to be replaced by the covariant derivative 
$\nabla^{(h)}_i$ on the momentum space $(\Gamma^+_m)_x$. 

\subsection{Relativistic damping force}

Recall that the damping force arises from the interaction between the Brownian 
particle and the heat reservoir, but only accounts for a portion of the total 
interaction, neglecting thermal fluctuations. We can directly measure the 
damping force when the thermal fluctuations can be ignored and establish 
a phenomenological model. It is reasonable to expect that the damping force 
should vanish if the Brownian particle comoves with the heat reservoir. 
Hence, the damping force can be regarded as an excitation of the 
relative velocity, with the damping coefficients serving as response 
factors. This idea was also explored in previous works \cite{debbasch1997relativistic, 
dunkel2005theory1} in the special relativistic context. Here we shall extend the construction 
to the general relativistic case and point out some crucial subtleties 
which need to be taken care of. 

In the relativistic context (be it special or general), the concept of velocity is 
replaced by proper velocity. However, the relative velocity {\em cannot} be defined
simply as the difference between two proper velocities, because the temporal  
component of the difference should not be considered as part of the relative velocity 
but rather as the energy difference. In order to have an appropriate definition for 
the relative velocity, one must project one of the two proper velocities onto the 
orthonormal direction of the other. Let $U^\mu$ be the velocity of the heat reservoir 
and $p^\mu$ be the proper momentum of the Brownian particle. 
One can associate with the Brownian particle an orthonormal projection tensor
\begin{align*}
\Delta^\mu{}_\nu(p):= 
\delta^\mu{}_\nu + \frac{p^\mu p_\nu}{m^2}
\end{align*}
which obeys
\[
\Delta^\mu{}_\nu(p) p^\nu=0.
\]
Then the relative velocity between the Brownian particle and the heat reservoir 
can be defined as $\Delta\ind{\mu}{\nu}(p)U^\nu$. This definition has two important 
features, i.e. (1) when the Brownian particle is comoving with the heat reservoir, 
the relative velocity vanishes; (2) the relative velocity is always normal to $p_\mu$, 
so that it is a vector in $T[(\Gamma^{+}_{m})_{x}]$. 

The relativistic damping force needs to have the following properties. First, it must 
contain the relative velocity as a factor, and a tensorial damping coefficient 
$\mathcal{K}^{\mu\nu}$ as another factor, i.e. 
\begin{align*}
\mathcal{F}^{\mu}_{\rm damp}=\mathcal{K}^{\mu\nu} \Delta_{\nu}{}^{\rho}(p)U_\rho. 
\end{align*}
Second, the damping force needs to be a tangent vector of the momentum space
$(\Gamma^{+}_{m})_{x}$, i.e. $\mathcal{F}^{\mu} \in T[(\Gamma^{+}_{m})_{x}]$.
This latter requirement implies that $\mathcal{K}^{\mu\nu}$ must satisfy the relation 
\begin{align}
\mathcal{K}^{\mu\nu}(x,p)=\Delta\ind{\mu}{\alpha}(p)
\mathcal{K}^{\alpha\beta}(x,p) \Delta_{\beta}{}^\nu(p).
\label{DDK}
\end{align}
In the light of eq.\eqref{DDK} and the idempotent property of the projection tensor, 
the relativistic damping force can be simply rewritten as
\begin{align}
\mathcal{F}^{\mu}_{\rm damp}=\mathcal{K}^{\mu\nu}U_\nu. \label{damping_force}
\end{align}

The constraint condition \eqref{DDK} over the tensorial damping coefficient has 
a very simple special solution 
\[
\mathcal{K}^{\mu\nu}=\kappa(x,p) \Delta^{\mu\nu}(p),
\]
where $\kappa(x,p)$ is a scalar function on the SoMS $\Sigma_t$
and is referred to as the friction coefficient. 
This particular choice of damping coefficient corresponds to isotropic 
damping force. If $\kappa(x,p)$ is constant, then damping force will become homogeneous. 
Therefore, the isotropic homogeneous damping force can be written as
\begin{align*}
\mathcal{F}^{\mu}_{\rm damp}=\kappa\Delta\ind{\mu}{\nu}(p)U^{\nu}.
\end{align*}

Let $e\ind{\mu}{\hat i}$ be the spatial comoving frame covectors associated 
with the Brownian particle and $E\ind{\hat i}{\nu}$ be the dual vectors. Then 
the projection tensor $\Delta\ind{\mu}{\nu}(p)$ can be written as
\[
\Delta\ind{\mu}{\nu}(p) = e\ind{\mu}{\hat i}E\ind{\hat i}{\nu}.
\]
The isotropic homogeneous damping force can be re-expressed as
\begin{align*}
\mathcal{F}^{\mu}_{\rm damp}=\kappa e\ind{\mu}{\hat i}E\ind{\hat i}{\nu}U^{\nu}
={\kappa U^{\hat{i}}}e\ind{\mu}{\hat i} ,
\end{align*}
or more concisely as 
\begin{align*}
	\mathcal{F}^{\hat{i}}_{\rm damp}=\kappa U^{\hat{i}},
\end{align*}
where $\mathcal{F}^{\hat{i}}_{\rm damp}= \mathcal{F}^{\mu}_{\rm damp} E\ind{\hat i}{\mu}$,
which represents the spatial components of the damping force under the comoving frame. 
This equation has the same form as the one given in 
\cite{dunkel2005theory1,dunkel2005theory2}. However, our expression \eqref{damping_force} 
for the damping force is more general and does not rely on a particular choice of frame basis.

\subsection{Covariant relativistic Langevin equation: a first attempt }

Although the intuitive form \eqref{intuitiveLE} of Langevin equation is 
mathematically unsounded, it is still inspiring while considering the extension of 
Langevin equation to generic spacetime manifolds. One can imagine that the relativistic 
Langevin equation should arise as the free geodesic motion of the Brownian particle 
perturbed by the extra damping and stochastic forces. Taking the proper time $\tau$
of the Brownian particle as evolution parameter, the geodesic equation can be 
rearranged in the form
\begin{align*}
\D x^\mu_\tau&=\frac{p^\mu_\tau}{m}\D \tau,\notag\\
\D p^\mu_\tau &=-\frac{1}{m} \Gamma^\mu{}_{\alpha\beta}p^\alpha_\tau p^\beta_\tau\D \tau,
\end{align*}
where $\Gamma^\mu{}_{\alpha\beta}$ is the usual Christoffel connection on the 
spacetime manifold $\mathcal M$.
Therefore, with the supplementation of Stratonovich's coupling with 
Gaussian noises, the additional stochastic force \eqref{additional} 
and the relativistic damping force 
\eqref{damping_force}, we can write down, as a first attempt, the following 
set of equations as candidate of relativistic Langevin equation,
\begin{align}
\D \tilde x^\mu_\tau&=\frac{\tilde p^\mu_\tau}{m}\D \tau,
\label{dxvsdtau}\\
\D \tilde p^\mu_\tau 
&=\left[\mathcal{R}^{\mu}{}_{\mathfrak{a}}\circ_S\D \tilde w_\tau^\mathfrak{a}
+\mathcal{F}^\mu_{\text{add}}\D \tau\right]
+\mathcal{K}^{\mu\nu}U_\nu\D \tau
-\frac{1}{m} \Gamma^\mu{}_{\alpha\beta}\tilde p^\alpha_\tau \tilde p^\beta_\tau\D \tau.
\label{langevin-relativity}
\end{align}
As previously mentioned, the Stratonovich's rule is the unique coupling rule which preserves 
the chain rule of calculus. Meanwhile, we have been very careful while introducing the 
damping and stochastic forces so that each of the first three force terms appearing 
on the right hand side of eq.\eqref{langevin-relativity} are tangent 
vectors of the momentum space $(\Gamma^+_m)_x$. With all these considerations combined 
together, eqs.\eqref{dxvsdtau}-\eqref{langevin-relativity} are guaranteed to be 
general covariant and have taken the damping and stochastic impacts from the 
heat reservoir into account. Moreover, since $\mathcal{R}^\mu{}_\mathfrak{a},\ 
\mathcal{F}^\mu_\add$ and $\mathcal{K}^{\mu\nu}$ are all tensorial objects 
on the future mass shell $(\Gamma^+_m)_x$, 
one can easily check that, provided the initial state is on the mass shell 
$(\Gamma^+_m)_x$, all future states evolving from 
eqs.\eqref{dxvsdtau}-\eqref{langevin-relativity} will remain on  $(\Gamma^+_m)_x$, 
because, for any $(\tilde x_\tau, \tilde p_\tau)$ obeying 
the mass shell condition 
\[
\tilde S_\tau=S(\tilde x_\tau, \tilde p_\tau) = 
g_{\mu\nu}(\tilde x_\tau) \,\tilde p^\mu_\tau \tilde p^\nu_\tau + m^2=0,
\]
we have
\begin{align}
\D \tilde S_\tau &= \pfrac{S}{x^\mu}\D \tilde x^\mu_\tau 
+\pfrac{S}{p^\mu}\D \tilde p^\mu_\tau \notag\\
&= \frac{1}{m}\partial_\mu g_{\alpha\beta}\,
\tilde p^\mu_\tau \tilde p^\alpha_\tau \tilde p^\beta_\tau \,\D \tau 
+ 2 g_{\mu\rho} \tilde p^\rho_\tau \(
\mathcal{R}^{\mu}{}_{\mathfrak{a}}\circ_S\D \tilde w_\tau^\mathfrak{a}
+\mathcal{F}^\mu \D \tau -\frac{1}{m} \Gamma^\mu{}_{\alpha\beta}
\tilde p^\alpha_\tau \tilde p^\beta_\tau\D \tau\) \notag \\
&= 2(\tilde p_\mu)_\tau 
\Big(\mathcal{R}^{\mu}{}_{\mathfrak{a}}\circ_S\D \tilde w_\tau^\mathfrak{a}
+\mathcal{F}^\mu \D \tau\Big) = 0,
\label{dStau}
\end{align}
where we have denoted $\mathcal{F}^\mu= \mathcal{F}^\mu_{\text{add}}
+\mathcal{K}^{\mu\nu}U_\nu$ for short. Eq.\eqref{dStau} implies that 
the $(d+1)$ components of $\tilde p^\mu$ are not all independent, and 
there is a redundancy contained in eq.\eqref{langevin-relativity}, 
which makes no harm due to the reason explained by 
eq.\eqref{relation-vector-on-mass-shell}.
In the end, it looks reasonable to consider 
eqs.\eqref{dxvsdtau}-\eqref{langevin-relativity} as a viable 
candidate for the relativistic Langevin equation in curved spacetime, 
and we will henceforth refer to this system of equations as LE$_\tau$. 

In the next section, we shall show that, from the phenomenological point of view, 
LE$_\tau$ still contains some issues which needs to be resolved. The crucial point lies 
in that, while considering the stochastic distribution of Brownian particles, one cannot 
rely on a comoving frame or observer. If we change to the view point of a regularly 
moving observer, the proper time $\tau$ of the Brownian particle itself will 
become a random variable and hence inappropriate to be used for parametrizing 
the stochastic motion of the system. Thus we need a reparametrization scheme 
to rewrite the relativistic Langevin equation 
in terms of the observer's proper time $t$ instead of $\tau$.

\section{Reparametrization}
\label{Reparametrization}

Recall that the configuration space $\mathcal S_t$ is inherently connected with a 
concrete choice of observer and is defined as the level set $t(x)=t$. 
Therefore, $\partial_\mu t$ must be proportional to the 
unit normal covector $Z_\mu$ (i.e. the proper velocity of the chosen observer). 
Let
\[
\lambda := \sqrt{-g^{\mu\nu}\partial_\mu t \partial_\nu t}=|\nabla t|,
\]
we can write
\begin{align}
\partial_\mu t = -\lambda Z_\mu.
\label{ptZ}
\end{align}
Therefore, on the worldline of the Brownian particle, we have
\begin{align}
\D t=\partial_\mu t \D \tilde x^\mu
=-\lambda Z_\mu\D \tilde x^\mu=-\lambda Z_\mu \frac{\D \tilde x^\mu}{\D \tau}\D \tau
=-\lambda \frac{Z_\mu \tilde p^\mu}{m}\D \tau. 
\label{dtvsdtau}
\end{align}
Since $Z_\mu \tilde p^\mu<0$, the last equality explains the sign convention 
that appeared in eq.\eqref{ptZ}. Let 
\begin{align}
\gamma(\tilde x,\tilde p):=-\frac{\lambda Z_\mu \tilde p^\mu}{m},
\label{vdef}
\end{align}
the relation \eqref{dtvsdtau} can be rewritten as: 
\begin{align}\label{relationoftime}
\D t=\gamma(\tilde x,\tilde p)\D \tau.
\end{align}
$\gamma(\tilde x,\tilde p)$ plays the role of a local Lorentz factor.
Since $\tilde x^\mu, \tilde p^\mu$ are both random, the regularity of the prescribed 
observer implies that the proper time $\tau$ of the Brownian particle becomes 
essentially a random variable. This poses a serious challenge to understanding 
eqs.\eqref{dxvsdtau}-\eqref{langevin-relativity} as the relativistic 
Langevin equation, because 
Langevin equation requires a regular evolution parameter. 

In spite of the challenge just mentioned, we still wish to make some sense of 
eqs.\eqref{dxvsdtau}-\eqref{langevin-relativity} 
and try to find a resolution of the problem that 
we encountered. For this purpose, we temporarily adopt a comoving description 
for the Brownian particle but nevertheless let Alice  
be bound together with the coordinate system, 
so that the coordinate time $x^0$ equals the proper time $t$ of Alice. 
Let us stress that binding the observer with the coordinate system 
is not an absolutely necessary step, but it indeed simplifies the following
discussions about the probability distributions. 
In this description, $\tau$ appears to be a regular variable, but then the spacetime
position $\tilde x^\mu$ (which contains the observer's proper time as a component) and 
momentum $\tilde p^i$ will become random variables 
depending on $\tau$. Due to the mass shell condition, there is no need to include 
$\tilde p^0$ in the set of micro state variables. 

Unlike regular variables, random variables do not have a definite value, but rather 
a probability distribution. Thus LE$_\tau$ provides insight into the evolution 
of the probability distribution of the random variables involved. The reason that 
LE$_\tau$ can provide a probability distribution relies on the fact that 
Stratonovich's coupling can be turned into Ito's coupling and that a stochastic 
differential equation with Ito's coupling can be viewed as a Markov process. 
Writing $X:=(x^\mu,p^i)$, the Markov process described by LE$_\tau$ provides 
the transition probability
\begin{align}\label{tranpro}
\Pr[\tilde X_{\tau+\D \tau}=X_{\tau+\D \tau}|\tilde X_{\tau}= X_{\tau}]
\end{align}
during an infinitesimal proper time interval $[\tau,\tau+\D \tau]$. Over a finite 
period of time, this will amount to the joint probability of the state of the Brownian 
particle and the observer's proper time at a given $\tau$,
\begin{align}
\Phi_\tau(t,x^i,p^i):=\Pr[\tilde{x}^0_\tau=t,\tilde x^i_\tau=x^i,\tilde p^i_\tau=p^i],
\label{Phitau}
\end{align}
\blue{which is normalized in the whole future mass shell $\Gamma_{m}^{+}$ 
under the measure provided by the volume element \eqref{volume-element}.} 
$\Phi_\tau(t,x^i,p^i)$ is connected with the transition probability \eqref{tranpro} via
\begin{align*}
\Phi_{\tau+\D\tau}(X)=\int\D Y \Pr[\tilde X_{\tau+\D \tau}
=X|\tilde X_{\tau}= Y]\Phi_\tau(Y).
\end{align*}

Although there are clear logic and corresponding mathematical tools to deal 
with the evolution equation of $\Phi_\tau$ from the comoving description of 
the Brownian particle, the probability function $\Phi_\tau$ is not a suitable 
object in statistical mechanics. Recall that the physically viable distribution in 
statistical mechanics must be a probability distribution on the SoMS,
while the definition of the SoMS, especially the configuration 
space $\mathcal S_t$, relies on the choice of observer. The problem of the 
probability distribution \eqref{Phitau} lies in that, for fixed $\tau$, different
realizations $x^\mu$ of $\tilde x^\mu$ do not necessarily fall in the 
same configuration space $\mathcal S_t$. 

Nevertheless, as we shall show in the next section by 
Monte Carlo simulation in the example case of $(1+1)$-dimensional 
Minkowski spacetime, we can still extract the physical probability distribution 
out of the result of eqs.\eqref{dxvsdtau}-\eqref{langevin-relativity}. 
The point lies in that 
one should not look at the distribution of the end points of each realization of the
Brownian motion after the fixed proper time period $\tau$. Rather, one should look at 
the distribution of the intersection points of the stochastic worldlines with the 
physical configuration space $\mathcal S_t$ (as will be shown in Fig.\ref{fig1}). 
The latter distribution is physical, but it looks challenging to obtain such a 
distribution by means of analytical analysis. 

A better way to obtain the physical probability distribution 
for the Brownian particle is to introduce a reparametrization for the 
Langevin equation, replacing the random parameter $\tau$ by the regular proper time 
$t$ of Alice, as will be discussed as follows. 
Let us mention that Dunkel {\em et al} \cite{dunkel2009time} has attempted to 
use reparametrization to make their special relativistic Langevin equation covariant. 
However, a general discussion for the necessity of reparametrization 
has not yet been persued. 
 
At the first sight, the reparametrization could be accomplished simply by substituting 
eq.\eqref{relationoftime} into eqs.\eqref{dxvsdtau}-\eqref{langevin-relativity}. 
However, things are not that simple. In order to get a physically viable Langevin equation, 
one need to ensure that the resulting equation should describe a Markov process
driven by a set of Wiener processes. To achieve this goal, 
we propose to first use discrete time nodes $t_n$ to treat the stochastic process 
as a Markov process, and then take the continuity limit. Here proper time $t$ 
needs {\em not} be identified with the coordinate time $x^0$. By defining a sequence 
of random variables using the discrete time nodes $t_n$, namely 
\begin{align}
\tilde\tau_n:=\tilde\tau_{t_n},\qquad
\tilde y^\mu_n:=\tilde x_{\tilde \tau_n}^\mu,\qquad
\tilde k^\mu_n:=\tilde p^\mu_{\tilde \tau_n},\qquad
\tilde Y_n:=\tilde X_{\tilde \tau_n},
\end{align}
we can calculate their discrete time differences,
\begin{align}
\D \tilde \tau_n&=\gamma^{-1}(\tilde Y_n)\D t_n, \label{tautn}\\
\D \tilde y^\mu_n &=\tilde x^\mu_{\tilde \tau_{n+1}}-\tilde x^\mu_{\tilde \tau_n} 
=\frac{\tilde k^\mu_n}{m}\gamma^{-1}(\tilde Y_n)\D t_n, \label{yn}\\
\D\tilde k_n^\mu &=\tilde p^\mu_{\tilde \tau_{n+1}}-\tilde p^\mu_{\tilde \tau_n} 
=F^\mu(\tilde Y_n) \gamma^{-1}(\tilde Y_n)\D t_n 
+ \mathcal R^\mu{}_\mathfrak{a}(\tilde Y_n)
\D \tilde w_{ \tilde \tau_n}^\mathfrak{a}. \label{fakelangevin}
\end{align}
In deriving eq.\eqref{fakelangevin}, we have changed the Stratonovich's rule 
in eq.\eqref{langevin-relativity} into Ito's rule before introducing the 
discretization, so that the total force $F^\mu$ reads
\begin{align*}
F^\mu=\frac{\delta^{\fa\fb}}{2}\mathcal R\ind{\nu}{\fa}
\frac{\partial}{\partial p^\nu}\mathcal R\ind{\mu}{\fb}
+\mathcal F_\add^\mu+\mathcal K^{\mu\nu}U_\nu
-\frac{1}{m}\Gamma\ind{\mu}{\alpha\beta}p^\alpha p^\beta.
\end{align*}

It is remarkable that, although eqs.\eqref{tautn}-\eqref{fakelangevin} appear to be 
complicated, they bear the enlightening feature that, at each time step, the 
increment of $(\tilde \tau_n, \tilde y^\mu_n, \tilde k_n^\mu)$ depend 
only on their values at the nearest proceeding time step. Therefore, 
we can understand these equations as describing a Markov process. However, 
these equations are not yet the sought-for reparametrized Langevin, because 
$\tilde w_{\tilde \tau_n}^\fa$ is no longer a Wiener process after 
the reparametrization. 

Fortunately, we can define a stochastic process
\begin{align}
\tilde W^\mathfrak{a}_n
:=\sum_{i=0}^n \gamma^{1/2}(\tilde Y_i ) \D \tilde w^\mathfrak{a}_{\tilde \tau_{i}},
\end{align}
whose increment at the $n$-th time step reads 
\begin{align*}
\D \tilde W^\mathfrak{a}_n
= \gamma^{1/2}(\tilde Y_n ) \D \tilde w^\mathfrak{a}_{\tilde \tau_{n}}.
\end{align*}
The conditional probability $\Pr[\D \tilde W^\mathfrak{a}_n
=\D W_n^\mathfrak{a}|\tilde Y_n=Y_n]$ can be easily calculated as
\begin{align*}
\Pr[\D \tilde W^\mathfrak{a}_n=\D W_n^\mathfrak{a}|\tilde Y_n=Y_n]
&=\frac{1}{\gamma^{\fd/2}(Y_n)}\frac{1}{(2\pi \D \tau_n)^{\fd/2}}
\exp \left [ -\frac{1}{2}\frac{\delta_{\mathfrak{a}\mathfrak{b}}
\D w_{ \tau_n}^\mathfrak{a} \D w_{ \tau_n}^\mathfrak{b}}{\D \tau_n} \right ]\notag\\
&=\frac{1}{(2\pi \D t_n)^{\fd/2}}
\exp \left [ -\frac{1}{2}\frac{\delta_{\mathfrak{a}\mathfrak{b}}
\D W_n^\mathfrak{a} \D W_n^\mathfrak{b}}{\D t_n} \right ].
\end{align*}
Since the above conditional probability is independent of the realization 
of $\tilde Y_n$, we can drop the condition,
\begin{align*}
\Pr[\D \tilde W^\mathfrak{a}_{n}=\D W^\mathfrak{a}_n]
&=\int \D Y_n \Pr[\D \tilde W^\mathfrak{a}_n=\D W^\mathfrak{a}_n|\tilde Y_n=Y_n] 
\Pr[\tilde Y_n=Y_n] \notag \\
&=\frac{1}{(2\pi \D t_n)^{\fd/2}}\exp \left [ 
-\frac{1}{2}\frac{\delta_{\mathfrak{a}\mathfrak{b}}\D W_n^\mathfrak{a} 
\D W_n^\mathfrak{b}}{\D t_n}\right ] \int \D Y_n  \Pr[\tilde Y_n=Y_n] \notag \\
&=\frac{1}{(2\pi \D t_n)^{\fd/2}}\exp \left [ -\frac{1}{2}
\frac{\delta_{\mathfrak{a}\mathfrak{b}}
\D W_n^\mathfrak{a} \D W_n^\mathfrak{b}}{\D t_n} \right ].
\end{align*}
Thus, in the continuum limit $\D t_n \to \D t$, $\tilde W^\mathfrak{a}_n$ becomes 
a standard Wiener process $\tilde W_t$ with variance $\D t$.
In the end, we obtain the following stochastic differential equations
as the continuum limit of eqs.\eqref{yn} and \eqref{fakelangevin},
\begin{align}
\D \tilde y^\mu_t&=\frac{\tilde k^\mu_t}{m}\gamma^{-1}\D t, \label{yt}\\
\D \tilde k^\mu_t 
&=\left[\hat{\mathcal{R}}^{\mu}{}_{\mathfrak{a}}\circ_S\D\tilde W_t^\mathfrak{a}
+\hat{\mathcal{F}}^\mu_{\text{add}}\D t\right]
+\hat{\mathcal{K}}^{\mu\nu}U_\nu\D t 
-\frac{1}{m} \Gamma^\mu{}_{\alpha\beta}\tilde k^\alpha_t \tilde k^\beta_t 
\gamma^{-1} \D t,
\label{langevin-relativity-observer}
\end{align}
in which we introduced the following notations,
\begin{align*}
\hat{\mathcal R}\ind{\mu}{\fa}:=\gamma^{-1/2}\mathcal R\ind{\mu}{\fa},
\qquad
\hat{\mathcal{K}}^{\mu\nu}:=\gamma^{-1}{\mathcal{K}}^{\mu\nu},
\qquad
\hat{\mathcal{F}}_\add^\mu:=\gamma^{-1}\mathcal F^\mu_\add
-\frac{\delta^{\fa\fb}}{2}\mathcal R\ind{\mu}{\fa}\mathcal R\ind{j}{\fb}
(\gamma^{-1/2}\nabla^{(h)}_j \gamma^{-1/2}).
\end{align*}

Notice that we have changed the coupling rule once again into the Stratonovich's rule, 
with guarantees that the resulting equations \eqref{yt}-\eqref{langevin-relativity-observer} 
are manifestly general covariant. Moreover, after the reparametrization, 
eqs.\eqref{yt}-\eqref{langevin-relativity-observer} still describe a stochastic 
process driven by some Wiener noises, and are now parametrized by the regular 
evolution parameter $t$ instead of the random variable $\tau$. Therefore, 
eqs.\eqref{yt}-\eqref{langevin-relativity-observer}
fulfill all of our anticipations, and we will refer to this system of equations 
as LE$_t$.

Please be reminded that, although the observer's proper time $t$ needs not to 
be identical with the coordinate time $y^0$, they {\em can be made} identical 
by introducing the artificial choice for the observer which is at rest in the 
coordinate system. On such occasions, $y^0$ and $t$ 
should be treated as equal, and we need to check that the zeroth 
component of eq.\eqref{yt} represents an identity. According to eq.\eqref{ptZ}, 
when $y^0=t$, we have 
\[
Z_\mu = -\lambda^{-1} \partial_\mu t = -\lambda^{-1}\delta_\mu{}^0.
\]
Thus we have
\[
\gamma^{-1}= -\frac{m}{\lambda Z_\mu \tilde k^\mu_t}=\frac{m}{\tilde k^0_t}.
\]
Inserting this result into eq.\eqref{yt}, one sees that the zeroth component 
yields an identity $\D \tilde y^0_t =\D t$.

\section{Monte Carlo simulation in the Minkowski case}
\label{MonteCarlo}

As advocated in the last section, it is possible to extract reasonable information 
about the physical distribution on the SoMS from LE$_\tau$,  
although the evolution parameter $\tau$ itself is a random variable
from the observer's perspective. In this subsection, we
shall exemplify this possibility by studying the simple case of stochastic motion
of Brownian particles in $(1+1)$-dimensional Minkowski spacetime driven by
a single Wiener process and subjects to an isotropic homogeneous damping coefficients. 

To be more concrete, we use the orthonormal coordinates $x^\mu=(t,x)$ and
let $E:=p^0$ and $p:=p^1$, so that the mass shell 
condition becomes $E=\sqrt{p^2-m^2}$. For isotropic thermal perturbations, 
the stochastic amplitude should satisfy
\begin{align*}
\mathcal R^\mu \mathcal R^\nu=D\Delta^{\mu\nu}(p)=\frac{D}{m^2}
    \begin{bmatrix}
        p^2 & Ep\\
        Ep& E^2
    \end{bmatrix},
\end{align*}
where $D$ is the diffusion coefficient. It's obvious that the 
stochastic amplitude should read
\begin{align*}
    \mathcal R^\mu= \frac{\sqrt{D}}{m}
    \begin{bmatrix}
        p \\
        E
    \end{bmatrix}.
\end{align*}
If we put the observer and the heat reservoir at rest, i.e. $Z=U=\partial_t$, 
the coordinate time will be automatically the proper time of the observer, 
and the isotropic damping force should be
\begin{align*}
\mathcal F_{\text{damp}}^\mu=\kappa\Delta^{\mu\nu}(p)U_\nu=-\frac{\kappa}{m^2}
    \begin{bmatrix}
        p^2 \\
        Ep
    \end{bmatrix}.
\end{align*}
The above choice of observer implies $\gamma=E/m$.

Since the projection tensor $\Delta_{\mu\nu}(p)$ is simultaneously 
the metric of the momentum space $(\Gamma^+_m)_x$ with ``determinant''
\begin{align*}
    \det \Delta_{ij}= \Delta_{11} =\frac{m^2}{E^2},
\end{align*}
the additional stochastic force can be evaluated to be
\begin{align*}
\mathcal F^\mu_\add =\frac{1}{2}\mathcal R\ind{\mu}{}\nabla^{(h)}_i\mathcal R\ind{i}{}
= \frac{1}{2}\mathcal R^\mu \frac{1}{\sqrt{\Delta_{11}}}
\frac{\partial }{\partial p}\(\sqrt{\Delta_{11}} \frac{\sqrt{D}}{m}E\)=0.
\end{align*}

With the above preparation, we can now write down the two systems of 
Langevin equations with evolution parameters $\tau$ and $t$ as
\begin{align}
\D \tilde x_\tau=\frac{\tilde p_\tau}{m}\D \tau,\qquad 
\D \tilde p_\tau=\frac{\sqrt{D}}{m}E\circ_I\D \tilde w_\tau
+\frac{D\tilde p_\tau}{2m^2}\D\tau-\frac{\kappa E\tilde p_\tau}{m^2}\D\tau,
\label{langtau}
\end{align}
and
\begin{align}
\D\tilde y_t=\frac{\tilde k_t}{E}\D t,\qquad 
\D\tilde k_t=\sqrt{\frac{DE}{m}}\circ_I\D\tilde W_t
+\frac{D\tilde k_t}{2Em}\D t-\frac{\kappa}{m}\tilde k_t\D t.
\label{langt}
\end{align}
Since the observer is now bound together with the coordinate system, the temporal 
components of the Langevin equations become either trivial or redundant. Therefore, 
in eqs.\eqref{langtau} and \eqref{langt}, only the spatial components are presented.

We are now ready to make the numeric simulation based on the above two systems of
equations. The (initial) values of the simulation parameters are listed in
Tab.\ref{tab1}.

\begin{table}[h]
    \centering
    \begin{tabular}{cccc|cccc|cc}
        \toprule
        Parameters & $D$ & $\kappa$ &$m$  & $\tilde x|_{\tau=0}$ & $\tilde p|_{\tau=0}$
        & $\tilde y|_{t=0}$ & $\tilde k|_{t=0}$ & $\D t$ & $\D\tau$\\
        \midrule
        (Initial) values  & 1.0 & 1.0 & 1.0 & 0 & 0 & 0 & 0 & 0.02 & 0.02\\
        \bottomrule
     \end{tabular}
     \caption{The (initial) values of the simulation parameters}
     \label{tab1}
\end{table}

\begin{figure}[h]
\centering
\includegraphics[scale=.8]{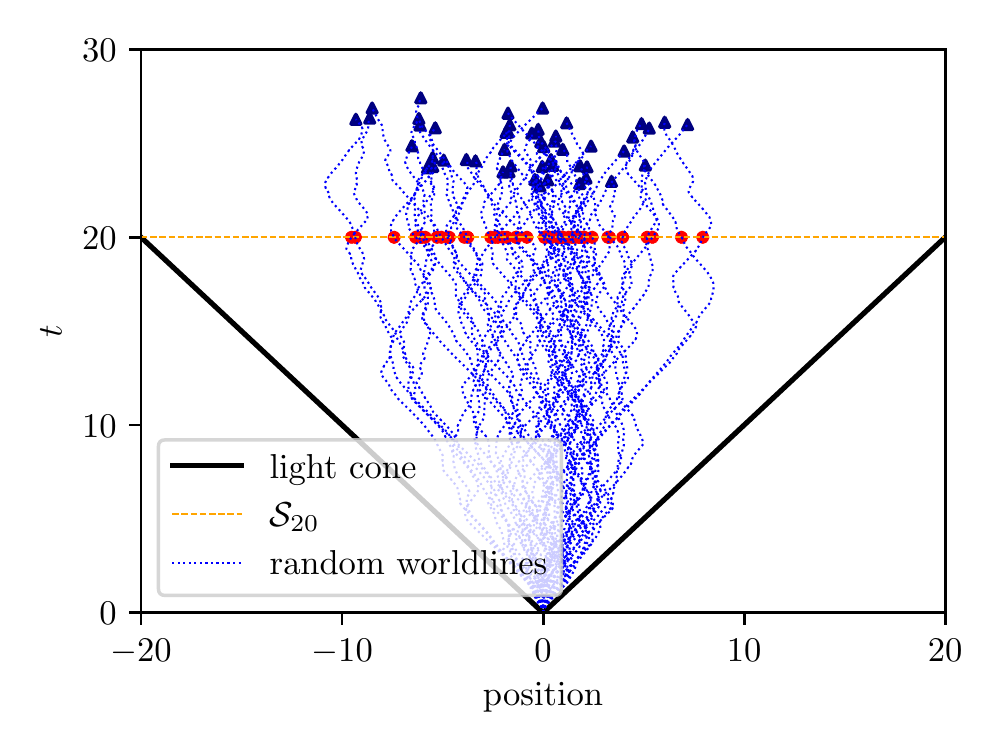}
\includegraphics[scale=.8]{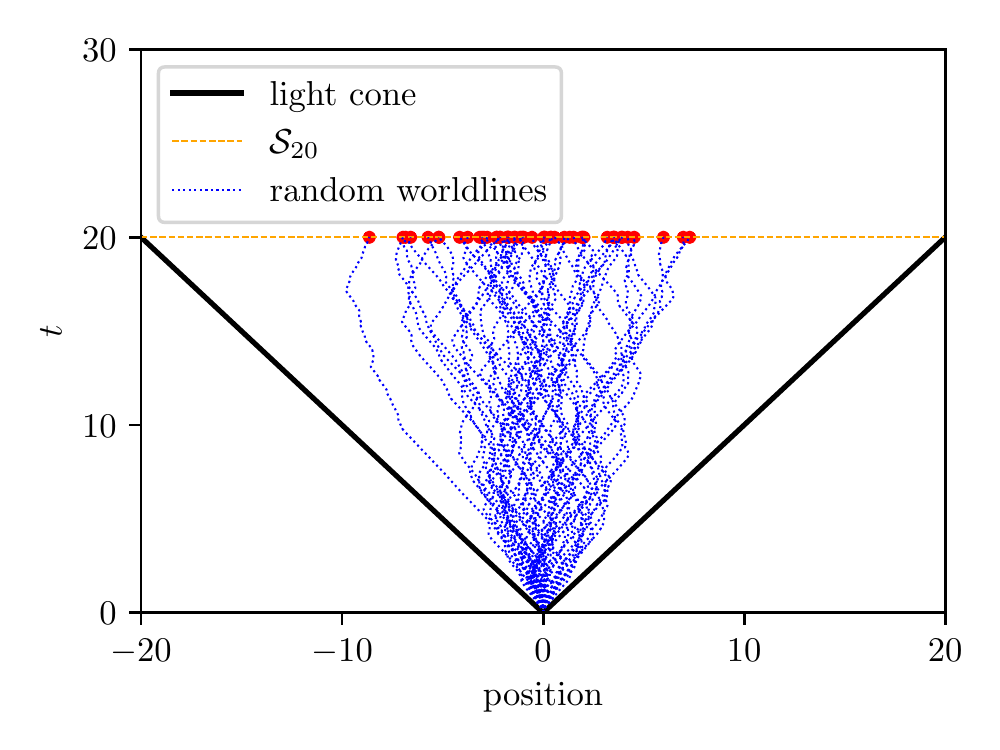}
\caption{Random worldlines generated by eq.\eqref{langtau} (left) 
and eq.\eqref{langt} (right)}
\label{fig1}
\end{figure}

The left picture in Fig.\ref{fig1} depicts 50 random worldlines generated by 
eq.\eqref{langtau} after a 
fixed ``evolution time'' $\tau=20$. The end point of each random worldline is marked by 
a solid triangle, and the horizontal line at $t=20$ represents the configuration space 
$\mathcal S_{20}$. We can see that all worldlines fall strictly in the future lightcone 
of the initial event $(t,x)=(0,0)$, and the end points of different random worldlines 
do not fall in the same configuration space. Nevertheless, we can extract the intersection 
point of each worldline with the configuration space $\mathcal S_{20}$ (marked
with round dots) and try to identify their distribution. 

The right picture in Fig.\ref{fig1} depicts 50 random worldlines generated 
by eq.\eqref{langt} after the
fixed evolution time $t=20$. Since $t$ is the regular evolution parameter, 
the end points of all worldlines automatically fall in the same configuration space
$\mathcal S_{20}$ and are marked with round dots. 
This gives an intuitive illustration for the power of 
the reparametrization introduced in the last section. One can feel how similarly the 
round points in both pictures in Fig.\ref{fig1} are distributed.  

\begin{figure}[h]
\centering
\includegraphics[scale=0.6]{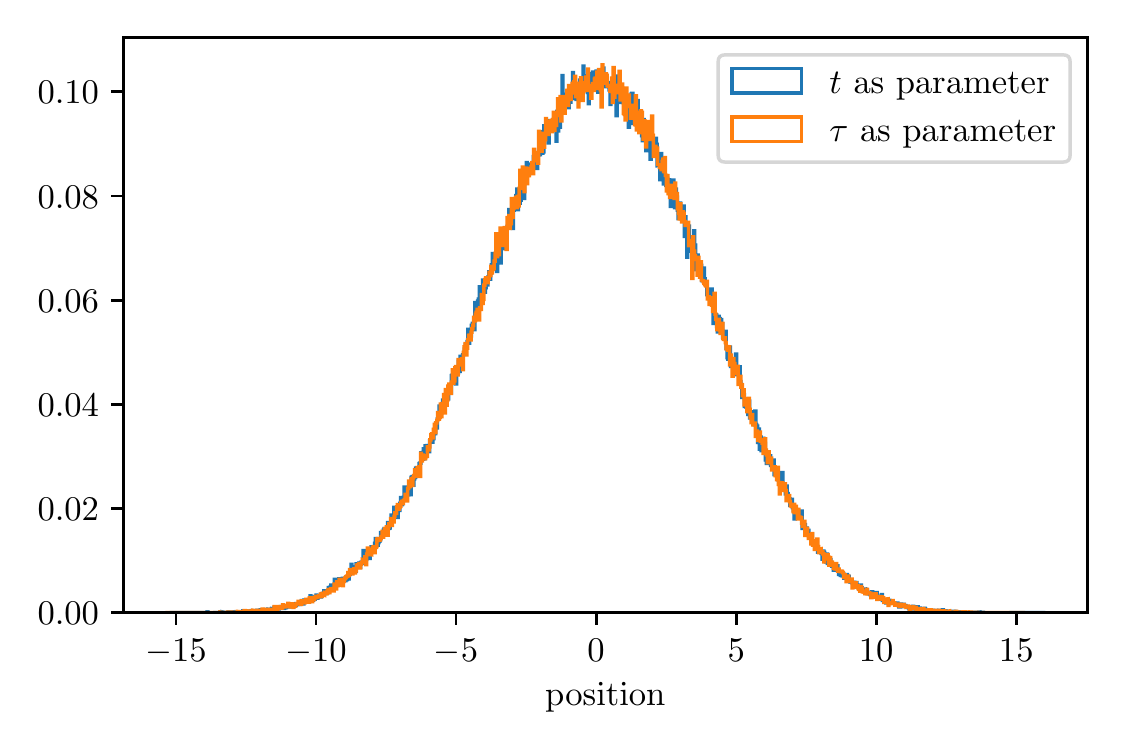}
\includegraphics[scale=0.6]{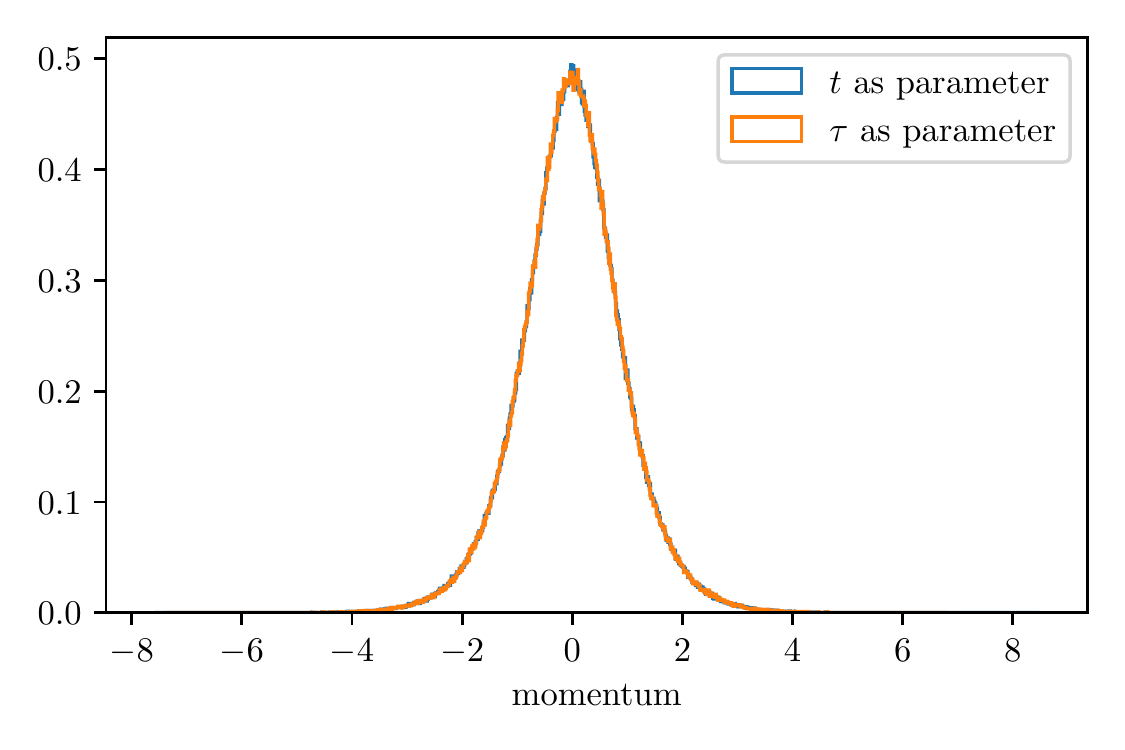}
\caption{The distributions of Brownian particles in configuration space 
$\mathcal S_{20}$ (left) and momentum space (right) from the two systems of equations 
\eqref{langtau} and \eqref{langt} }
\label{fig2}
\end{figure}

With a little more efforts, we have generated $10^6$ random phase trajectories 
using both eq.\eqref{langtau} and eq.\eqref{langt}
and collected the data for the insection points of the random worldlines  
with the configuration space $\mathcal S_{20}$ in the case of 
eq.\eqref{langtau}. Using the data thus collected, we can depict separately the
configuration space and momentum space distributions of the Brownian particles 
and make comparisons between the results that follow from
eq.\eqref{langtau} and eq.\eqref{langt}. As can be seen in Fig.\ref{fig2}, 
the results from eq.\eqref{langtau} and eq.\eqref{langt} are almost identical.

\begin{figure}[h]
\centering
\includegraphics[scale=0.5]{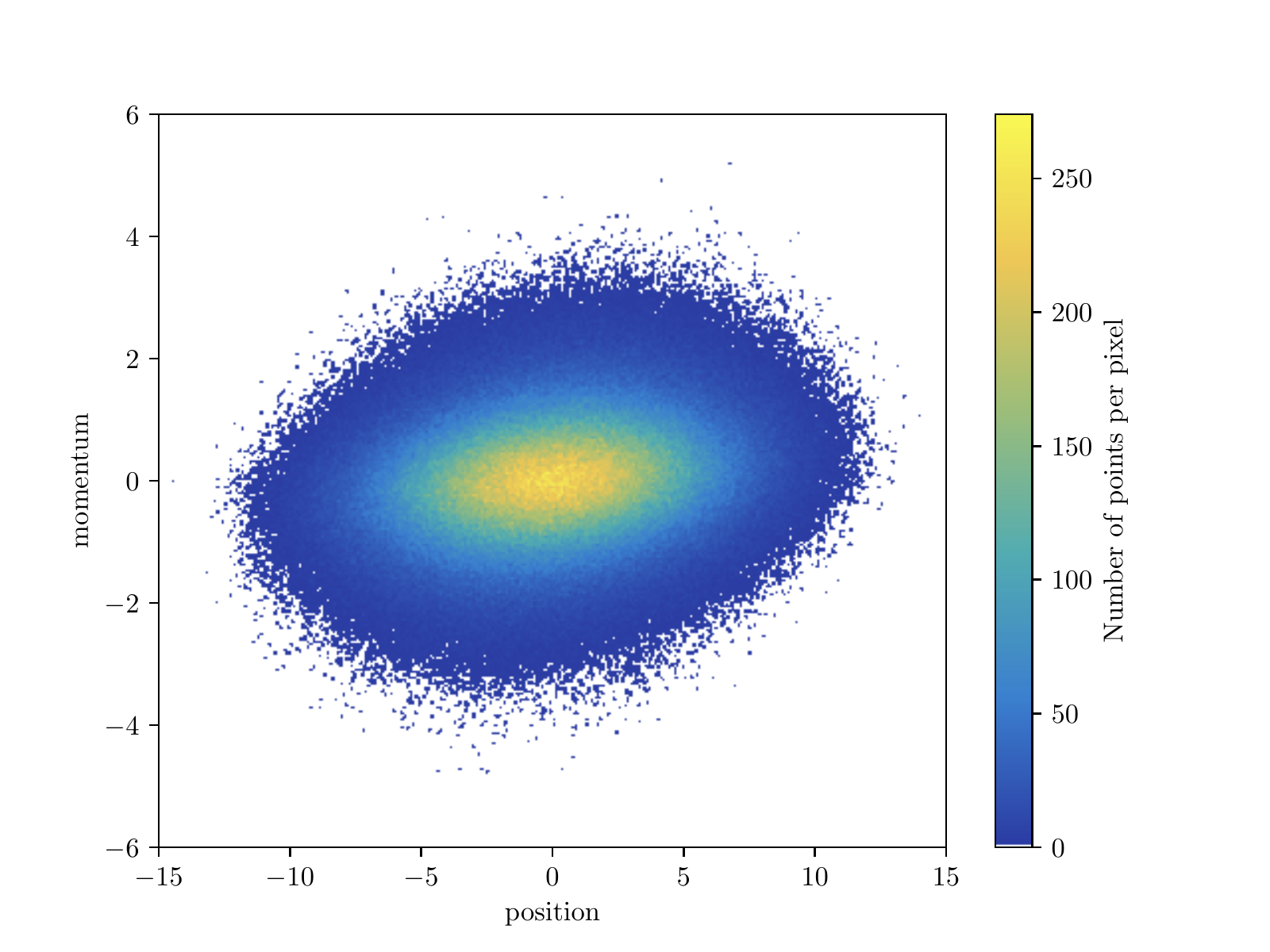}
\includegraphics[scale=0.5]{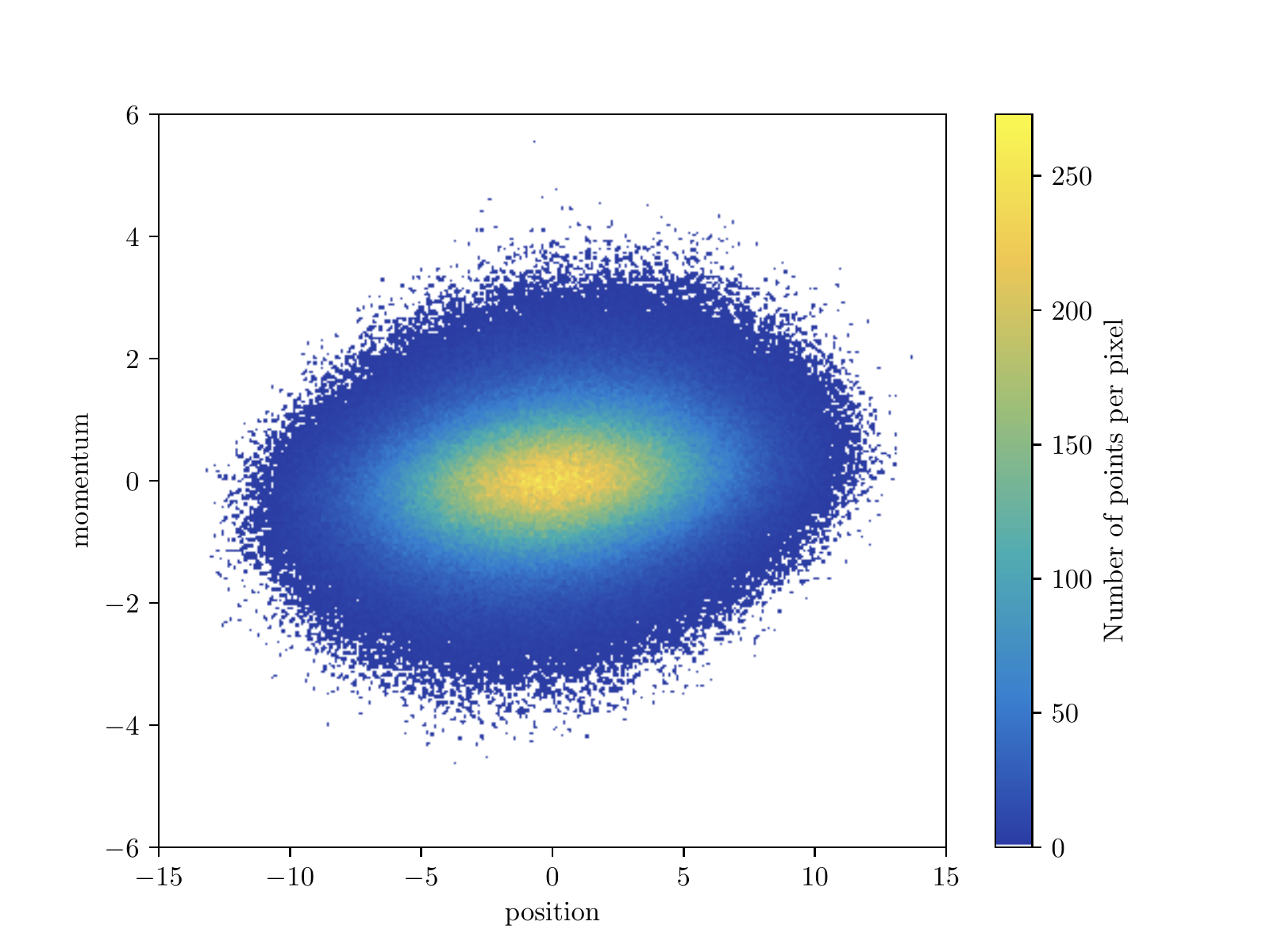}
\caption{Joint probability distributions in positions and momenta at $t=20$: 
The left picture arises from eq.\eqref{langtau}, and the right one arises from 
eq.\eqref{langt}. The distributions presented in Fig.\ref{fig2} correspond to 
vertical and horizontal projections of these joint distributions.}
\label{fig4}
\end{figure}

We also generated the joint distributions in positions 
and momenta from both eqs.\eqref{langtau} and \eqref{langt} at $t=20$. 
The results are presented in Fig.\ref{fig4}. We can hardly find any 
differences between the two pictures.

As a more serious comparison between the distributions generated by 
eqs.\eqref{langtau} and \eqref{langt}, the 
Pearson $\chi^2$-test is utilized with the assistance of Wolfram Language 
to determine whether the distributions were indeed identical. 
The resulting P-values were found to be $0.774$ for the distributions 
presented in the left plots of Fig.\ref{fig2}, 
$0.967$ for distributions presented in the right plots of Fig.\ref{fig2}, 
and $0.972$ for the two distributions presented in Fig.\ref{fig4}. 
These results provide more solid evidence for the expectation that the 
distributions generated by the two systems of equations \eqref{langtau} 
and \eqref{langt} are identical.

\section{Concluding remarks}
\label{Conclusion}

We have thus formulated two different versions of the relativistic Langevin equation,
i.e. LE$_\tau$ and LE$_t$ in a generic curved spacetime 
background, which are both manifestly general covariant. The two versions differ from 
each other in that LE$_\tau$ takes the proper time $\tau$
of the Brownian particle as evolution parameter, while LE$_t$ 
takes the proper time $t$ of the prescribed observer Alice as evolution parameter.

The importance of the prescribed regularly moving observer is stressed throughout 
the analysis, especially while clarifying the SoMS of 
the Brownian particle and interpreting the probability distributions of 
the Brownian particle. It is argued that, in order to get the physical 
probability distribution, LE$_t$ is more preferable than LE$_\tau$. We also discussed the 
conditions which the relativistic damping coefficients need to obey,
and clarified the concept of relative velocity in the relativistic context.

We also demonstrated, by means of Monte Carlo simulation 
in the particular example case of Brownian motion in $(1+1)$-dimensional Minkowski 
spacetime, that although LE$_\tau$ contains some conceptual issues, 
it is indeed possible to extract physically reasonable probability distributions 
from it. However, since the Brownian particles after a fixed 
proper time $\tau$ do not fall in the same configuration space, it would be 
more difficult to obtain the physical probability distributions from LE$_\tau$. 

This work is the first of our attempts for a systematic study of general relativistic
stochastic mechanics. In a forthcoming work, we will proceed to formulate the 
corresponding Fokker-Planck equations and discuss the physical consequences. 
In particular, the general relativistic variant of Einstein relation will be 
considered, and the relationship between different probability distributions 
will be clarified.

Before ending this paper, let us mention that there is another complementary 
approach, i.e. the 2-jet bundle approach \cite{armstrong2016coordinate,
armstrong2018intrinsic} using Ito's formalism, 
for describing the covariant Brownian motion \cite{meyer2006differential,
schwartz1984semimartingales,emery2012stochastic,kuipers2021stochastic1,
kuipers2021stochastic2}, see also \cite{ kuipers2023stochastic} for a more
recent review. Our formalism does not need to make use of 
the jet bundle, and the resulting equations are more in line with the original 
intuitive construction of Langevin. There is some other recent work \cite{Paraguassu} 
which focuses on the heat distribution in Minkowski spacetime, which has some 
overlap in research subjects with the present work.

\section*{Acknowledgement}

This work is supported by the National Natural Science Foundation of China under the grant
No. 12275138.

\section*{Data Availability Statement} 

All data used in this research was generated from eqs.\eqref{langtau} and \eqref{langt}  
by use of numeric programs written in C++ and python. The P-values given in Sec.5 
are calculated using Wolfram Language. All programs are available upon 
request. 

\section*{Declaration of competing interest}

The authors declare no competing interest.

\providecommand{\href}[2]{#2}\begingroup
\footnotesize\itemsep=0pt
\providecommand{\eprint}[2][]{\href{http://arxiv.org/abs/#2}{arXiv:#2}}


\begin{thebibliography}{10}

\bibitem{Juttner1911}
J\"uttner, F. ``Das Maxwellsche Gesetz der Geschwindigkeitsverteilung
in der Relativtheorie'',
\href{https://doi.org/10.1002/andp.19113390503}{\emph{Ann. der Phys.}
{339}, 856--882, 1911}.

\bibitem{degroot1980relativistic}
de~Groot, S.~R.; van Leeuwen, W.~A.; van Weert, Ch.~G.
{\em Relativistic Kinetic Theory: Principles and Applications.}
\newblock North-Holiand Publishing Company, 1980,
\newblock {ISBN: 9780444854537}.

\bibitem{cercignani2002relativistic}
Cercignani, C; Kremer, G.~M.
{\em The Relativistic Boltzmann Equation: Theory and Applications},
  volume~22.
\newblock \href{https://link.springer.com/book/10.1007/978-3-0348-8165-4}
{Springer Science \& Business Media, 2002. ISBN: 9783034881654}.

\bibitem{vereshchagin2017relativistic}
Vereshchagin, G.~V.;  Aksenov, A.~G.
\newblock {\em Relativistic kinetic theory: with applications in astrophysics
and cosmology}.
\href{https://www.cambridge.org/hk/academic/subjects/physics/astrophysics/relativistic-kinetic-theory-applications-astrophysics-and-cosmology?format=HB&isbn=9781107048225}{Cambridge University Press, 2017. ISBN: 9781107048225}.

\bibitem{acuna2022introduction}
Acu{\~n}a-C{\'a}rdenas, R.~O.; Gabarrete, C.; Sarbach, O.
``An introduction to the relativistic kinetic theory on curved
spacetimes.''
\href{https://doi.org/10.1007/s10714-022-02908-5}
{{\em Gen. Rel. and Grav.} 54(3): 1--120, 2022}.[\eprint{2106.09235}].

\bibitem{debbasch1997relativistic}
Debbasch, F.; Mallick, K.; Rivet, J.~P.
``Relativistic Ornstein--Uhlenbeck process,''
\href{https://doi.org/10.1023/B:JOSS.0000015180.16261.53}{{\em J. Stat. Phys.} 
88(3): 945--966, 1997}.

\bibitem{debbasch2004diffusion}
Debbasch, F.
``A diffusion process in curved spacetime,''
\href{https://doi.org/10.1063/1.1755860}{{\em J. Math. Phys.} 45(7): 2744--2760, 2004}.

\bibitem{dunkel2005theory1}
Dunkel, J.; H{\"a}nggi, P.
``Theory of relativistic Brownian motion: the (1+1)-dimensional case.''
\href{https://doi.org/10.1103/PhysRevE.71.016124}
{{\em Phys. Rev. E} 71(1): 016124, 2005}. [\eprint{cond-mat/0411011}].

\bibitem{dunkel2005theory2}
Dunkel, J.; H{\"a}nggi, P.
``Theory of relativistic Brownian motion: The (1+ 3)-dimensional case,''
\href{https://doi.org/10.1103/PhysRevE.72.036106}
{{\em Phys. Rev. E} 72(3): 036106, 2005}. [\eprint{cond-mat/0505532}].

\bibitem{fingerle2007relativistic}
Fingerle, A.
``Relativistic fluctuation theorems,''
\href{https://doi.org/10.1016/j.crhy.2007.05.015}{{\em C. R. Phys.} 
8(5-6): 696--713, 2007}. [\eprint{0705.3115}].

\bibitem{franchi2007relativistic}
Franchi, J.; Le~Jan, Y.
``Relativistic diffusions and Schwarzschild geometry,''
\href{https://doi.org/10.1002/cpa.20140}{{\em Commun. Pure Appl. Math.} 
60(2): 187--251, 2007}. [\eprint{math/0410485}].

\bibitem{dunkel2009relativistic}
Dunkel, J.; H{\"a}nggi, P.
``Relativistic Brownian motion,''
\href{https://doi.org/10.1016/j.physrep.2008.12.001}
{{\em Phys. Rep.} 471(1): 1--73, 2009}. [\eprint{0812.1996}].

\bibitem{dunkel2009time}
Dunkel, J.; H{\"a}nggi, P.; Weber, S.
``Time parameters and Lorentz transformations of relativistic stochastic processes,''
\href{https://doi.org/10.1103/PhysRevE.79.010101}{{\em Phys. Rev. E} 79(1): 010101, 2009}. 
[\eprint{0812.0466}].

\bibitem{herrmann2009diffusion}
Herrmann, J.
``Diffusion in the special theory of relativity,''
\href{https://doi.org/10.1103/PhysRevE.80.051110}
{{\em Phys. Rev. E} 80(5): 051110, 2009}.[\eprint{0903.0751}].

\bibitem{herrmann2010diffusion}
Herrmann, J.
``Diffusion in the general theory of relativity,''
\href{https://doi.org/10.1103/PhysRevD.82.024026}
{{\em Phys. Rev. D} 82(2): 024026, 2010}. [\eprint{1003.3753}].

\bibitem{haba2010relativistic}
Haba, Z.
``Relativistic diffusion with friction on a pseudo-Riemannian manifold,''
\href{https://doi.org/10.1088/0264-9381/27/9/095021}{{\em Class. Quant. Grav.} 
27(9): 095021, 2010}. [\eprint{0909.2880}].

\bibitem{ding2020covariant}
Ding, M.; Tu, Z.; Xing, X.
``Covariant formulation of nonlinear Langevin theory with
multiplicative gaussian white noises,''
\href{https://doi.org/10.1103/PhysRevResearch.2.033381}
{{\em Phys. Rev. Res.} 2(3): 033381, 2020}. [\eprint{2007.16131}].

\bibitem{ding2022covariant}
Ding, M.; Xing, X.
``Covariant nonequilibrium thermodynamics from Ito-Langevin dynamics,''
\href{https://doi.org/10.1103/PhysRevResearch.4.033247}{{\em Phys. Rev. Res.} 
4(3): 033247, 2022}.[\eprint{2105.14534v2}].

\bibitem{einstein1905neue}
Einstein, A.
{\em Eine neue bestimmung der molek{\"u}ldimensionen}, 
\href{https://doi.org/10.3929/ethz-a-000565688}{PhD thesis, ETH Zurich, 1905}.

\bibitem{einstein1905molekularkinetischen}
Einstein, A.
``{\"U}ber die von der molekularkinetischen theorie der w{\"a}rme
geforderte bewegung von in ruhenden fl{\"u}ssigkeiten suspendierten teilchen,''
\href{https://doi.org/10.1002/andp.19053220806}{{\em Ann. der Phys.} 4, 1905}.

\bibitem{smoluchowski1906kinetischen}
Smoluchowski, M.
``Zur kinetischen theorie der Brownschen molekular bewegung und der
  suspensionen,''
\href{https://doi.org/10.1002/andp.19063261405}{{\em Ann. der Phys.} 21: 756--780, 1906}.

\bibitem{langevin1908theorie}
Langevin, P.
``Sur la th{\'e}orie du mouvement Brownien,''
{\em C. R. Acad. Sci.} 146: 530--533, 1908.

\bibitem{ford1965statistical}
Ford, G.~W.; Kac, M.; Mazur, P.
``Statistical mechanics of assemblies of coupled oscillators,''
 \href{https://doi.org/10.1063/1.1704304}{{\em J. Math. Phys.} 6(4): 504--515, 1965}.

\bibitem{mori1965transport}
Mori, H.
``Transport, collective motion, and Brownian motion,''
\href{https://doi.org/10.1143/PTP.33.423}{{\em Prog. Theor. Phys.} 33(3): 423--455, 1965}.

\bibitem{zwanzig1973nonlinear}
Zwanzig, R.
``Nonlinear generalized Langevin equations,''
\href{https://doi.org/10.1007/BF01008729}{{\em J. Stat. Phys.} 9(3): 215--220, 1973}.

\bibitem{sekimoto1998langevin}
Sekimoto, K.
``Langevin equation and thermodynamics,''
\href{https://doi.org/10.1143/PTPS.130.17}{{\em Prog. Theor. Phys. Supp.} 130: 17--27, 1998}.

\bibitem{sekimoto2010stochastic}
Sekimoto, K.
{\em Stochastic energetics, volume 799.}
\href{https://link.springer.com/book/10.1007/978-3-642-05411-2}{Springer, 2010. 
ISBN: 9783642054112}.

\bibitem{sarbach2014geo}
Sarbach, O.; Zannias, T.
``The geometry of the tangent bundle and the relativistic kinetic
theory of gases,''
\href{https://doi.org/10.1088/0264-9381/31/8/085013}{{\em Class. Quant. Grav.} 
31(8): 085013, 2014}. [\eprint{1309.2036}].

\bibitem{gardiner1985handbook}
Gardiner, C.~W. 
{\em Handbook of stochastic methods for physics, chemistry and the 
natural sciences, volume~3.}
\href{https://link.springer.com/book/9783540208822}{Springer Berlin, 
1985. ISBN: 9783540208822}.

\bibitem{hsu2002stochastic}
Hsu, E.~P.
{\em Stochastic analysis on manifolds, Number~38.}
\href{https://bookstore.ams.org/gsm-38}
{American Mathematical Society, 2002. ISBN: 9780821808023}.

\bibitem{armstrong2016coordinate}
Armstrong, J.; Brigo, D.
``Coordinate-free stochastic differential equations as jets,'' 
[\eprint{1602.03931}].

\bibitem{armstrong2018intrinsic}
Armstrong, J.; Brigo, D.
``Intrinsic stochastic differential equations as jets,''
\href{https://doi.org/10.1098/rspa.2017.0559}{{\em Proceedings of the Royal Society A: 
Mathematical, Physical and
Engineering Sciences}, 474 (2210): 20170559, 2018}.

\bibitem{klimontovich1994nonlinear}
Klimontovich, Y.~L.
``Nonlinear Brownian motion,''
\href{https://doi.org/10.1070/PU1994v037n08ABEH000038}{{\em Phys-Usp}, 37(8): 737, 1994}.

\bibitem{meyer2006differential}
Meyer, P.~A. 
``A differential geometric formalism for the It{\^o} calculus''
\href{https://doi.org/10.1007/BFb0088719}
{{\em Stochastic Integrals: Proceedings of the LMS Durham Symposium July 7-17, 1980}, Springer Berlin, Heidelberg, 1981. ISBN: 9783540106906}.

\bibitem{schwartz1984semimartingales}
Schwartz, L.
{\em Semimartingales and their stochastic calculus on manifolds.}
{Gaetan Morin Editeur Ltee, 1984. ISBN: 9782760606609}.

\bibitem{emery2012stochastic}
{\'E}mery, M.
{\em Stochastic calculus in manifolds.}
\href{https://link.springer.com/book/10.1007/978-3-642-75051-9}
{Springer Science \& Business Media, 2012. ISBN: 9783642750519}.

\bibitem{kuipers2021stochastic1}
Kuipers, F.
``Stochastic quantization on lorentzian manifolds,''
\href{https://doi.org/10.1007/JHEP05(2021)028}{{\em J. High Energ. Phys.}, 
2021(5): 1--51, 2021}. [\eprint{2101.12552}].

\bibitem{kuipers2021stochastic2}
Kuipers, F.
``Stochastic quantization of relativistic theories,''
\href{https://doi.org/10.1063/5.0057720}{{\em J. Math. Phys.}, 62(12): 122301, 2021}. 
[\eprint{2103.02501}].

\bibitem{kuipers2023stochastic}
Kuipers, F.
{\em Stochastic Mechanics: The Unification of Quantum Mechanics with Brownian Motion.}
\href{https://doi.org/10.1007/978-3-031-31448-3}
{Springer Cham, 2023. ISBN: 9783031314476}.

\bibitem{Paraguassu}
Paraguassu, P.~V.; Morgado, W.~A.~M. 
``Heat distribution of relativistic Brownian motion,'' 
\href{https://doi.org/10.1140/ep jb/s10051-021-00214-8}
{{\em Eur. Phys. J. B} 94: 197, 2021.}

\end{thebibliography}

\end{document}